\documentclass[preprint,proceedings]{rmaa}

\newcommand{\D}{\discretionary{}{}{}}

\SetYear{2013}
\SetConfTitle{LARIM XIV Latin American Regional IAU Meeting}

\title{Solar astrometry with Rio Astrolabe \& Heliometer}

\author{C. Sigismondi\altaffilmark{1,2}, S. C. Boscardin\altaffilmark{2}, A. H. Andrei\altaffilmark{2}, E. Reis-Neto\altaffilmark{3}, J. L. Penna\altaffilmark{2}, V. A. D'\'Avila\altaffilmark{2}}

\altaffiltext{1}{ICRA, Rome, Italy www.icra.it/solar sigismondi\D{}@icra.\D{}it}
\altaffiltext{2}{Observat\'orio Nacional \& $^{3}$MAST, Rio de Janeiro, Brazil} 

\suppressfulladdresses

\listofauthors{C. Sigismondi, S. C. Boscardin, A. H. Andrei, E. Reis-Neto, J. L. Penna, V. A. D'Avila}
\indexauthor{Sigismondi, C., Boscardin, S. C., Andrei, A. H., Reis-Neto, E.,Penna,  J. L., D'Avila,  V. A.}

\addkeyword{Astronomical instrumentation, methods and techniques}
\addkeyword{Instrumentation: high angular resolution}
\addkeyword{Astrometry}
\addkeyword{Sun: fundamental parameters}

\begin{document}
\maketitle 

\boldabstract{Abstract: Monitoring the micro-variations of the solar diameter helps to better understand local and secular trends of solar activity and Earth climate. The instant measurements with the Reflecting Heliometer of Observat{\'o}rio Nacional in Rio de Janeiro (RHRJ) have minimized optical and thermal distortion, statistically reducing air turbulence effects down to 0.01 arcsec.  Contrarily to satellites RHRJ has unlimited lifetime, and it bridges and extends the measures made with drift-scan timings across altitude circles with 0.1 arcsec rms with Astrolabes. The Astrolabe in Rio (ARJ) operated from 1998 to 2009 to measure the solar diameter and the detected variations have statistical significance.}

~~~~{\bf Heliometer:} The 11 cm parabolic mirror of RHRJ is splitted on its half, forming an appropriate fixed angle $\beta$. On the focal plane two images of the Sun are formed and the distance $d$ between the limbs is linked to the solar diameter $D_{\odot}$ by the equation $d=\beta-D_{\odot}$. A neutral mylar filter of transmittance $10^{-5}$ screens the Sun.
	The images are free from chromatic aberrations and thermal focal variations are minimized by the mirrors made in CCZ HS and the tube in carbon fiber, both with linear thermal expansion $\alpha \le 2 \times 10^{-8}/^{o}C$. 
RHRJ performs $\ge 1000$ measurements per day at all heliolatitudes [1]. 
Vertical and horizontal differential refractions to the solar image are computed: they are minimum when the solar center approaches the zenith as in Rio. Therefore RHRJ is the ideal instrument to monitor the solar diameter from the ground, and to bridge satellites and astrolabes historical series of data. 

~~~~{\bf Astrolabes:} They have been the standard instruments for solar astrometry since the 1970s. A 10 cm horizontal telescope (ARJ is refracting like all Danjon astrolabes, DORAYSOL is reflecting) receives the solar light from a isosceles prism, with its base in vertical position. The upper face of the prism reflects the light onto the objective, and the lower one onto a mercury mirror which defines the ideal horizontal surface and then sends the light to the objective. The instrument remains fixed during the transit (drift) of the Sun through the altitude circle defined by the prism angle, and the contact times of direct \& reflected images are extrapolated from the video. The prism angle of ARJ can be changed manually by regulating a spring, in order to enlarge the number of measurements to max 30 per day. For solar astrometry reference both ARJ and RHRJ have effective openings of $\sim4\times7$ cm.
With RHRJ we proved that the Astrolabe glass filters used to select the image wavelength within 100 nm act as a {\sl sub-arcsecond} wedge producing a superimposed secondary image systematically shifted of the same wedge angle. Hence all Astrolabes, Solar Disk Sextant and Picard data can be plotted to the same reference value, even if the original averages are separated by $\ge 3 \sigma$ one from another: these shifts are compatible with the tiny wedge angles, measurable so accurately only with the heliometer[2].

~~~~{\bf Conclusions:} The fluctuations of solar diameter during the last four decades, as measured with Astrolabes and ARJ [figure], have to be considered statistically significant, as the eclipses data and the planetary transits data confirm.
Picard satellite observations of metrologic quality are covering the last 3 years, coincident with an unusual low solar activity similar with the Dalton minimum of 1810.
The micro variations of the solar diameter claimed by some authors in this last decade can depend on solar activity.
The measurements of solar diameter from ground as RHRJ and Picard-Sol extend satellite measurements to decades, needed to fully understand the secular variations of the solar activity well known as responsible of major climate changes.
\begin{figure}[!t]
  \includegraphics[width=\columnwidth]{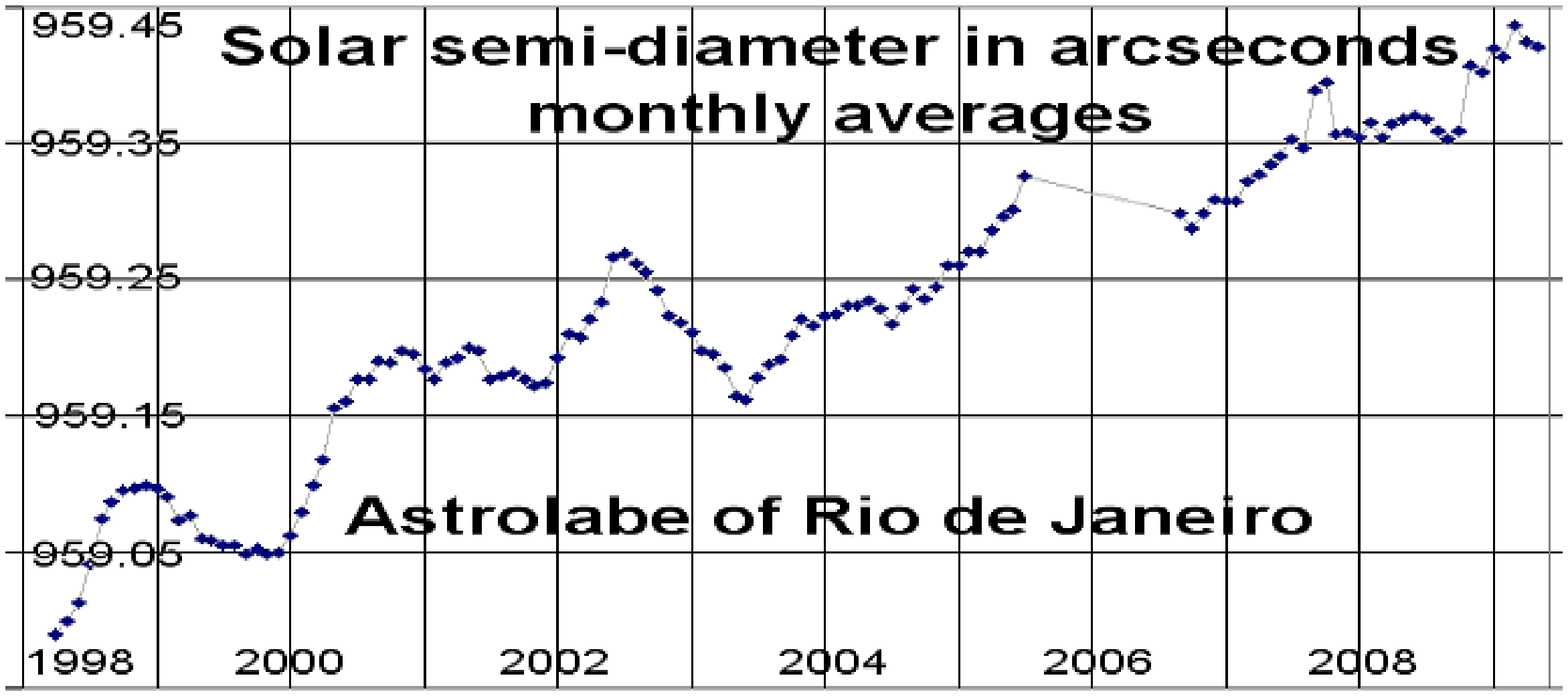}
\end{figure}

\end{document}